# RESULTS OF THE FURTHER ANALYSIS OF TIEN-SHAN ARRAY DATA ON THE ENERGY SPECTRUM OF PRIMARY COSMIC RAYS AT $2 \times 10^{13} - 3 \times 10^{17}$ eV

©2019  E.N. Gudkova, N.M. Nesterova

*P.N. Lebedev Physical Institute of the Russian Academy of Sciences*

## ABSTRACT

The energy spectrum of primary cosmic rays at energies between $2 \cdot 10^{13} - 3 \cdot 10^{17}$ eV is presented according to data from the Tien Shan array on the basis of the detection of the number of electrons in extensive air showers. In the energy range $5 \cdot 10^{15} - 3 \cdot 10^{17}$ eV, the spectrum was obtained by means of the HADRON array and was extended to the region of lower energies from $2 \cdot 10^{13}$ eV on the basis of the results of an individual experiment. The changes in the slope of the spectrum in the energy range of $10^{16} - 3 \cdot 10^{17}$ eV and a feature of this spectrum at about $10^{17}$ eV are analyzed in detail and are described. The spectrum in question is compared with the results obtained at some other arrays.

## INTRODUCTION

On the basis of experimental data from the Tien Shan array (690 g·cm$^{-2}$) [1–4], the energy spectrum of primary cosmic rays (PCR) was obtained over a wide range of primary energies, $E_0 = 2 \cdot 10^{13} - 3 \cdot 10^{17}$ eV; these results were based on the spectrum of extensive air showers (EAS) with respect to the number of electrons, $Ne$, which was determined from the readings of a system of scintillation detectors with allowance for their calibration accomplished with the aid of gas-discharge counters. In the energy range of $E_0 = 2 \cdot 10^{13} - 10^{15}$ eV, the spectrum was obtained in an individual experiment intended for detecting low-energy EAS [5, 6].

## PROCESSING OF EXPERIMENTAL DATA AND ANALYSIS OF RESULTS

The PCR energy spectrum was calculated on the basis of the EAS spectrum with respect to the number of electrons, $Ne$. A transition from $Ne$ to the primary energy $E_0$ is accomplished by simulating generated EAS on the basis of the MQ1 model proposed by Dunaevsky and his coauthors [7].

In just the same way as in [8] (2017), but in difference to what was done in [4] (1995), the $E_0$ spectrum from the data obtained at the HADRON array was rescaled from the spectrum of showers with respect to the number of electrons, $Ne$, by applying a new, refined, algorithm for determining the parameter $S$ of the lateral distribution of electrons (EAS age).

Showers in the zenith-angle range of $\theta < 30^0$ were selected here in data processing. On the basis of a simulation of the detection and selection of EAS in the array, the distance $R$ from the array center within which the shower-detection efficiency is 100% was determined for showers of different energy. The results are the following: $R < 10$ m for $E_0 = 2 \cdot 10^{13} - 10^{15}$ eV, $R < 20$ m for $E_0 = 3 \cdot 10^{15} - 10^{16}$ eV, and $R < 55$ m for $E_0 = 10^{16} - 3 \cdot 10^{17}$ eV.

In contrast to [3, 4], we performed the present analysis of the HADRON data by employing those taken within the last period of exploitation of the array, when the shower detection was the most reliable. Moreover, some criteria for event rejection in the original EAS database were disabled. This led, among other things, to an increase in the PCR intensity for $E_0 > 5 \cdot 10^{15}$ eV in relation to the results in [8].

In what concerns the chemical composition of PCR, our measurements at the HADRON array [8] revealed a change in the fraction of different nuclei as the energy grows in the range of $E_0 = 10^{15} - 3 \cdot 10^{17}$ eV. One observes a substantial increase in the fraction of "heavy" nuclei in PCR as the energy $E_0$ grows from $5 \cdot 10^{15}$ eV to $E_0 \sim 10^{17}$ eV: the fraction of EAS characterized by a gently sloping lateral distribution of electrons (large value of the age $S$) becomes higher by several orders of magnitude. The results of other experiments also suggest an increase in this fraction. In the region of $E_0 = 10^{16} - 3 \cdot 10^{17}$ eV, one also observes some increase in the number of EAS characterized by small values of $S$ and initiated by "light" nuclei [8].

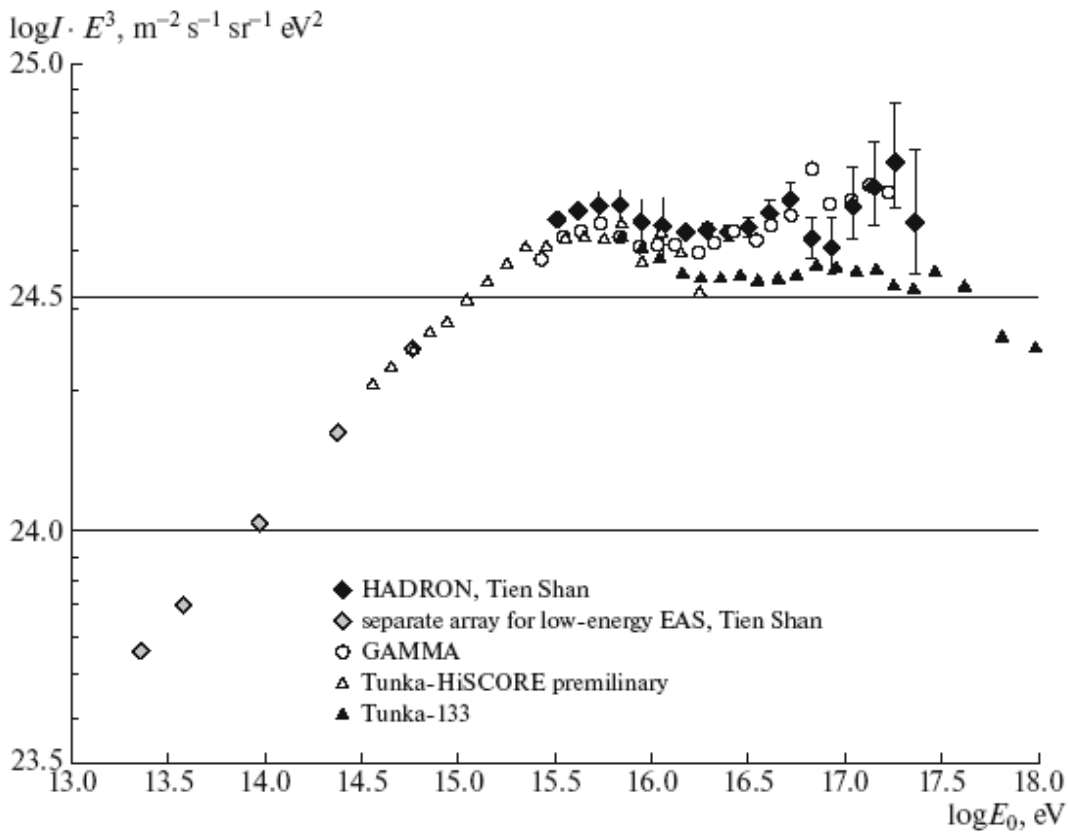

**Fig. 1.** Differential energy spectrum of primary cosmic rays, $I \cdot E_0^3$, according to data from the following arrays: ◆ – HADRON (Tien Shan array), ◇ – separate array for extensive air showers (EAS) of low energy (Tien Shan array), ○ – GAMMA(Aragatz), △ – Tunka-HiSCORE Preliminary (Baikal region), and ▲ – Tunka-133 (Baikal region).

It is noteworthy that the dependence of the relationship between $N_e$ and $E_0$ on the nuclear PCR composition is weaker at the mountain level than at altitudes near sea level.

## COMPARISON OF THE RESULTS WITH DATA FROM OTHER EXPERIMENTS

For the sake of comparison with the HADRON spectrum, Fig. 1 shows the spectra obtained at the GAMMA array [9] in the energy range of $E_0 = 10^{15} - 3\cdot10^{17}$ eV and the Tunka array [10, 11] in the energy range of $E_0 = 3\cdot10^{14} - 10^{18}$ eV. In addition, Fig. 2 gives the spectra from the HiRes [12], Ice-TOP [13], TALE [14], and Yakutsk [15] arrays for various energy ranges lying between $10^{15}$ and $10^{18}$ eV. Figure 3 illustrates a comparison of our spectrum with four versions of the spectrum obtained by the KASCADE-Grande Collaboration [16] on the basis of the SIBYLL, EPOS, and EPOS-LHC models. Here, one can clearly see a substantial dependence of the result on the model used. In the HADRON spectrum, the first change of the inclination occurs in the region of the known "knee" at $E_0 \sim 3\cdot10^{15}$ eV. In the region of energies above $2\cdot10^{16}$ eV (see Figs. 1–3, which show the spectrum in units of $I \cdot E_0^3$, where $I$ is the PCR intensity), the spectrum becomes more gently sloping. At still higher energies in the range of $E_0 = (5 - 8)\cdot10^{16}$ eV, we can see a rise followed by a descent. In the region around $E_0 \sim 10^{17}$ eV, the spectrum displays a jump. This jump remains present in various data samples that differ in the sample size, step width, boundaries of the intervals under analysis, etc., and this gives more grounds to believe in its statistical significance.

In addition to HADRON, similar changes in the spectrum in the range of $E_0 = 10^{15} - 5\cdot10^{17}$ eV were observed at the majority of other arrays. These changes include a "knee" at $E_0 \sim 3\cdot10^{15}$ eV, the region of a more gently sloping behavior from $E_0 \sim 2\cdot10^{16}$ eV, and a rise of different magnitude followed by a descent in the region from $6\cdot10^{16}$ to $10^{17}$ eV.

It is noteworthy that a jump in the spectrum at $E_0 \sim 10^{17}$ eV was found at the GAMMA array, which is located at approximately the same altitude above sea level as the Tien Shan array.

## CONCLUSIONS

The differential energy spectrum of primary cosmic rays was obtained at the Tien Shan array over a wide energy range, $E_0 = 2\cdot10^{13} - 3\cdot10^{17}$ eV. This spectrum is shown, along with the spectra from some other arrays, in Fig. 1 for $E_0$ from $2\cdot10^{13}$ eV and in Figs. 2 and 3 for $E_0$ from $10^{15}$ eV. In the energy range under study, the slope of the spectrum undergoes a number of changes described in the section Comparison of the Results with Data from Other Experiments. In the range of $E_0 = (5-8)\cdot10^{16}$ eV and in the region around $E_0 \sim 10^{17}$ eV, a leap appears in the spectrum.

A similar behavior of the spectrum was also recorded at other arrays and was described in a number of publications. In different experiments, the beginning and the magnitude of the rise in the spectrum at $E_0 \sim 10^{17}$ eV are different, which may be explained by the use of different models for rescaling the observed parameter of EAS to $E_0$ and by the difference of means for data analysis. For example, the SIBYLL, EPOS, EPOS-LHC, QGSJETII-04, and QGSJETII-02 models tested by the KASCADE Grande Collaboration [16] led to a difference in the intensity by a factor extending to two. Figure 3 illustrates a comparison of four of these spectra with our data.

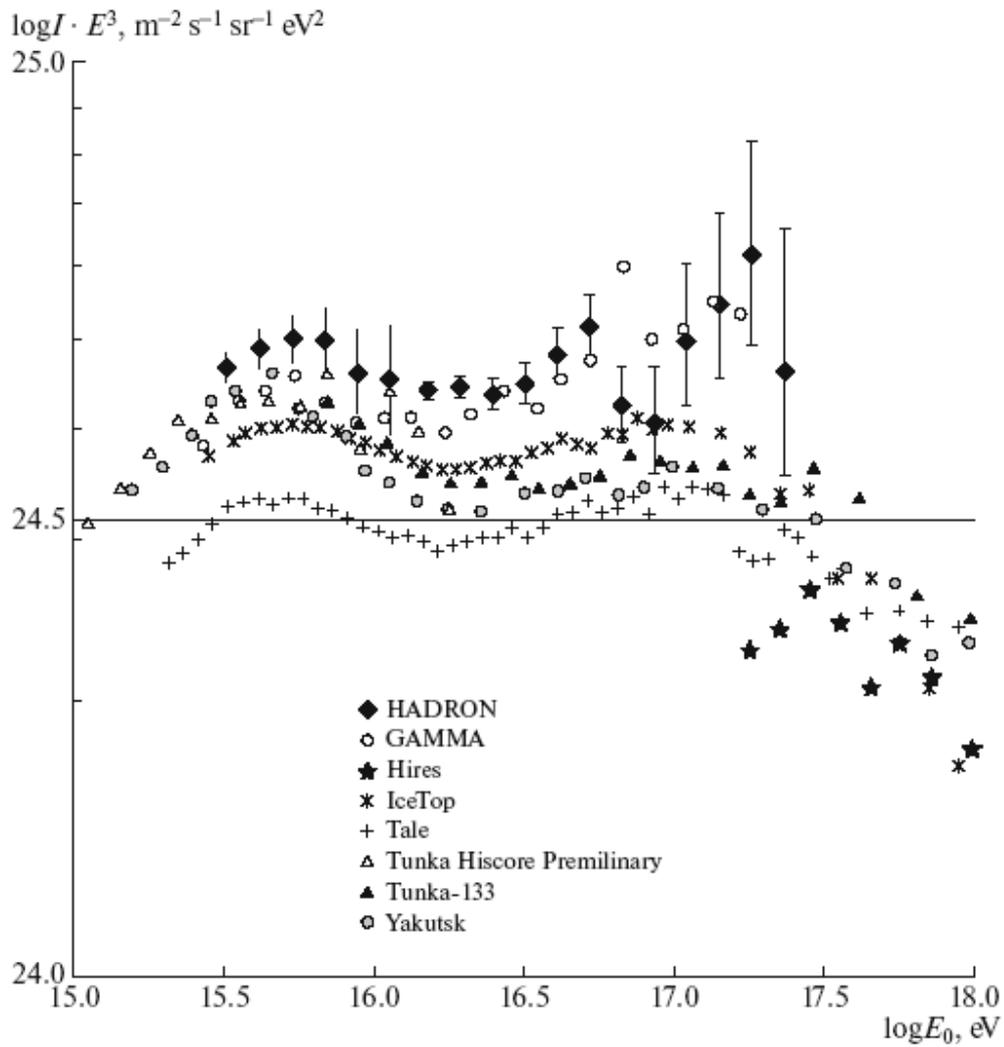

**Fig. 2.** Differential energy spectrum of primary cosmic rays, $I \cdot E_0^3$, according to data from the following arrays: ◆ –HADRON (Tien-Shan array), ○ – GAMMA, ★ – HiRes, ✳ – Ice-TOP, + – TALE, △ – Tunka-HiSCORE Preliminary, ▲ – Tunka-133, and ⬢ – Yakutsk.

The data obtained at the Tien Shan array [17] from the energy spectrum of hadrons with energies above 1 TeV on the growth of the inelastic interaction cross section for PCR protons in air in the energy range extending up to 10 PEV comply with the QGSJETII-04 model. In the following, this model can be used to rescale the $N_e$ spectrum to the PCR energy spectrum.

In the region around $E_0 \sim 10^{17}$ eV, it is necessary to assume a change in the fraction of different nuclei in the chemical composition [8] (especially an increase in the fraction of heavy nuclei) in relation to what we have at lower energies.

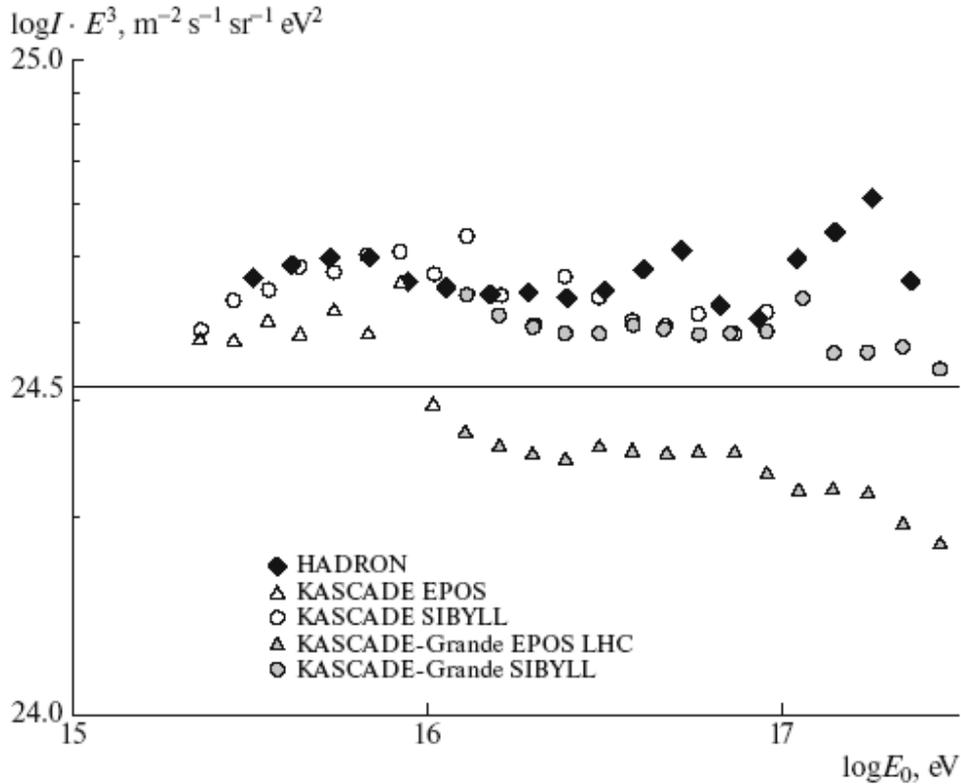

**Fig. 3.** Differential energy spectrum of primary cosmic rays, $I \cdot E_0^3$, according to data from the following arrays: ◆ – HADRON (Tien Shan array), △ – KASCADE EPOS, ○ – KASCADE SIBYLL, ▲ – KASCADE-Grande EPOS LHC, and ● – KASCADE-Grande SIBYLL.

Since changes in the slope of the spectrum in the energy range of $E_0 = 3 \cdot 10^{15} - 10^{18}$ eV were also observed at many arrays other than the HADRON array, including GAMMA, HiRes II, Ice-TOP, KASCADE Grande, TALE, TIBET, and Yakutsk, these changes may be a manifestation of new physics processes in PCR and call for a detailed investigation and interpretation. In order to explain them, several hypotheses were put forth in the literature (see [9, 16, 18, 19], for example).

## ACKNOWLEDGMENTS

We are grateful to our numerous colleagues involved in the studies reported in [1–4] and to all those who took part in creating the Tien-Shan array, performing respective measurements, composing the database, and developing algorithms for processing and analysis of the results.

## REFERENCES

1. D. S. Adamov, V. V. Arabkin, K. V. Barkalov, N. G. Vildanov, A. G. Dubovy, A. D. Erlykin, B. B. Kadyrsisov, S. K. Machavariani, R. A. Nam, N. M. Nesterova, S. I. Nikolsky, V. P. Pavluchenko, V. P. Stavrev, K. V. Cherdyntseva, A. P. Chubenko, and S. B. Shaulov, in *Proceedings of the 20th International Cosmic Ray Conference, Moscow, USSR, 1987*, HE. 6, p. 144.
2. D. S. Adamov, V. V. Arabkin, N. G. Vil'danov, L. I. Vil'danova, P. A. Dyatlov, N. S. Konovalova, S.K.Machavariani, N. M.Nesterova, S. I. Nikol'skiĭ, V. V. Piskal', S. A. Polishchuk, K. V. Cherdyntseva, A. P. Chubenko, A. L. Shchepetov, S. B. Shaulov, Izv. Akad. Nauk SSSR, Ser. fiz. **55**, 703 (1991).

3. L. I. Vildanova, N.M. Nesterova, and A. P. Chubenko, Phys. At. Nucl. **57**, 2145 (1994).
4. N. M. Nesterova, A. P. Chubenko, P. A. Djatlov, and L. I. Vildanova, in *Proceedings of the 24th International Cosmic Ray Conference, Roma, Italy, 1995,* Vol. 2, p. 748.
5. V. S. Aseikin, S. K.Machavariani, S. I. Nikolsky, and E. I. Tukish, in *Proceedings of the 18th International Cosmic Ray Conference, Bangalore, India, 1983,* Vol. 8, p. 71.
6. S. K. Machavariani, N. M. Nesterova, S. I. Nikolsky, V. A. Romakhin, and E. I. Tukish, in *Proceedings of the 17th International Cosmic Ray Conference, Paris, France, 1981,* Vol. 6, p. 193.
7. A. M. Dunaevsky et al., AIP Conf. Proc. **276**, 136 (1995).
8. E. N. Gudkova, N. M. Nesterova, N. M. Nikolskaya, and V. P. Pavlyuchenko, Bull. Russ. Acad. Sci.: Phys. **81**, 457 (2017).
9. R. M. Martirosov, A. P. Garyaka, H. S. Vardanyan, A. D. Erlykin, N. M. Nikolskaya, Y. A. Gallant, L. W. Jones, H. A. Babayan, et al., J. Phys.: Conf. Ser. **409**, 12081 (2012); arXiv: 1201.0235.
10. V. V. Prosin, S. F. Berezhnev, N. M. Budnev, M. Br Ë uckner, A. Chiavassa, O. A. Chvalaev, A. V. Dyachok, S. N. Epimakhov, A. V. Gafarov, O. A. Gress, T. I. Gress, D. Horns, N. N. Kalmykov, N. I. Karpov, S. N. Kiryuhin, E. N. Konstantinov, et al., EPJWeb Conf. **99**, 04002 (2015).
11. V. V. Prosin, S. F. Berezhnev, N. M. Budnev, A. Chiavassa, O. A. Chvalaev, A. V. Dyachok, S. N. Epimakhov, O. A. Gress, T. I. Gress, N. N. Kalmykov, N. I. Karpov, E. N. Konstantinov, A. V. Korobchenko, E. E. Korosteleva, L. A. Kuzmichev, et al., EPJ Web Conf. **121**, 03004 (2016).
12. P. Sokolsky and G. B. Thomson, J. Phys. G **34**, R401 (2007); arXiv: 0706.1248 [astro-ph].
13. Ice Cube–Gen2 Collab., PoS (ICRC2015) 694.
14. R. U. Abbasi, M. Abe, T. Abu-Zayyad, M. Allen, R. Azuma, E. Barcikowski, J. W. Belz, D. R. Bergman, S. A. Blake, R. Cady, B. G. Cheon, J. Chiba, M. Chikawa, A. Di Matteo, T. Fujii, K. Fujita, et al., arXiv: 1803.01288v1 [astro-ph.HE].
15. S. Knurenko, I. Petrov, Z. Petrov, and I. Sleptsov, PoS (ICRC2015) 252.
16. M. Bertaina, W. D. Apel, J. C. Arteaga-Vel r azquez, K. Bekk, J. Bl Ë umer, H. Bozdog, I. M. Brancus, E. Cantoni, A. Chiavassa, F. Cossavella, K. Daumiller, V. de Souza, F. di Pierro, P. Doll, R. Engel, D. Fuhrmann, et al., PoS (ICRC2015) 359.
17. N.M. Nesterova, EPJWeb Conf. **145**, 19001 (2017).
18. A. D. Erlykin and A. W. Wolfendale, J. Phys. G **23**, 979 (1997).
19. S. B. Shaulov and S. P. Bezshapov, EPJ Web Conf. **52**, 04010 (2013).